\documentclass[twocolumn,aps,superscriptaddress,showpacs]{revtex4}

\usepackage{amssymb}
\usepackage{amsmath}
\usepackage{graphicx}
\usepackage[normalem]{ulem}
\usepackage[dvips]{color}

\usepackage[normalem]{ulem} 
\usepackage[dvips]{color} 
\renewcommand\sout{\bgroup \color{red} \ULdepth=-.5ex \ULset}

\begin{document}

\title{Disentangling effects of collision geometry and symmetry energy in U+U collisions}
\author{Jun Xu}\email{Jun.Xu@tamuc.edu}
\affiliation{Department of Physics and Astronomy, Texas A$\&$M
University-Commerce, Commerce, TX 75429-3011, USA}
\author{Zachary Martinot}\email{zmartinot@gmail.com}
\affiliation{Department of Physics and Astronomy, Texas A$\&$M
University-Commerce, Commerce, TX 75429-3011, USA}
\affiliation{Department of Physics, University of Washington,
Seattle, WA 98195, USA}
\author{Bao-An Li}\email{Bao-An.Li@tamuc.edu}
\affiliation{Department of Physics and Astronomy, Texas A$\&$M
University-Commerce, Commerce, TX 75429-3011, USA}
\affiliation{Department of Applied Physics, Xi'an Jiao Tong
University, Xi'an 710049, China}

\begin{abstract}
Effects of the collision geometry on experimental observables that
are known to be sensitive to the high-density behavior of nuclear
symmetry energy are examined in U+U collisions at 0.52 GeV/nucleon
using an isospin- and momentum-dependent interaction within the
framework of IBUU transport model. It is found that the
neutron-proton differential flow in tip-tip collisions and the
difference of neutron and proton elliptic flow in body-body
collisions are more sensitive to the symmetry energy at
supra-saturation densities compared with collisions of spherical
nuclei of the same masses. In addition, the n/p ratio of
pre-equilibrium nucleons is found to be slightly more sensitive to
the symmetry energy in tip-tip collisions, and the collision
geometry affects the $\pi^-/\pi^+$ ratio significantly.
\end{abstract}

\pacs{25.70.-z, 
      24.10.Lx, 
      21.65.-f  
      }

\maketitle

\section{Introduction}\label{introduction}

Energetic heavy-ion collisions are the only way of producing
high-density nuclear matter in terrestrial laboratories. All heavy
nuclei are neutron-rich and many of them have a prolate or oblate
shape in their ground states. While the role of
deformation/orientation of nuclei in nuclear fusion/fission at low
energies has been studied extensively for a long time, it is only
during the past decade that we witnessed a surge of investigations
making good use of the deformation/orientation of heavy nuclei in
energetic nuclear reactions to probe interesting physics issues,
such as properties of the Quark-Gluon Plasma (QGP)
\cite{peter,ed99b,Kol00,Hei05,Nep07,Fil09,Hir11,Haq12} and chiral
magnetic effect \cite{Vol10} in relativistic heavy-ion collisions.
It is exciting to note that U+U experiments at 193 GeV/nucleon have
been done very recently by the STAR Collaboration at RHIC
\cite{Gag12}. While waiting for the experimental results, it is worth
mentioning that a very rich array of phenomena were predicted to
occur in U+U collisions due to different relative orientations of
the football-shaped uranium nuclei.  Among all possible
orientations, the tip-tip (with long axes head-on) and body-body
(with short axes head-on and long axes parallel) collisions are the
most interesting ones. In relativistic heavy-ion collisions, the
central body-body U+U collisions provides a better system to study
the properties of formed QGP compared with Pb+Pb or Au+Au collisions
of similar eccentricity~\cite{Kol00,Hei05}. Furthermore, the
particle multiplicity, initial eccentricity, and final collective
flow will be affected by the collision
geometry~\cite{Nep07,Fil09,Hir11,Haq12}. At beam energies of a few
hundred MeV/nucleon, tip-tip collisions normally have a larger
stopping power and a longer high-density phase compared to body-body
collisions~\cite{Luo07,Cao10}. It has also been found that the
multiplicities of pions and free nucleons are different in tip-tip
collisions and body-body collisions~\cite{Li00a,Luo07,Cao10}. In
addition, the transverse flow, which is an observable sensitive to
the nuclear matter equation of state (EOS), is much larger in
non-central tip-tip collisions than in non-central body-body
collisions~\cite{Luo07}. Moreover, the elliptic flow ($v_2$), which
is another messenger about nucleon-nucleon interactions in dense
matter, is different in tip-tip and body-body collisions as
well~\cite{Li00a,Luo07,Cao10}. While it is very challenging to
select events of special orientations, such as tip-tip and body-body
in U+U collisions, several promising triggers have been proposed in
the literature~\cite{Nep07,Wu08} and tested positively in model
simulations.

Generally speaking, heavy nuclei are all neutron-rich especially in
their surface areas. Moreover, protons and neutrons may have
different deformations. There are considerable interests in the
heavy-ion reaction community to carry out U+U collisions at
intermediate energies to probe the density dependence of nuclear
symmetry energy $E_{sym}(\rho)$ at 1-3 times normal nuclear matter
density. For instance, U+U experiments at 0.52 GeV/nucleon have
already been planned at the external target facility of the cooling
storage ring at Lanzhou/China~\cite{Xu,Xiao}.  The $E_{sym}(\rho)$
at supra-saturation density is among the most uncertain properties
of dense neutron-rich matter and has many important ramifications in
astrophysics~\cite{Ste05,Lat07,LCK08}. Especially, the
$E_{sym}(\rho)$ at 1-3 times normal nuclear matter density plays the
most important role in determining the radii of neutron
stars~\cite{Lat01}. Does the U+U collisions really provide a better
opportunity to probe the EOS of dense neutron-rich matter than
reactions with spherical nuclei of similar masses? Suppose one can
indeed trigger on collisions with specific orientations in U+U
collisions, which orientation, tip-tip or body-body, may be a better
choice to investigate the $E_{sym}(\rho)$ at supra-saturation
densities? For a given orientation, what are the sensitive probes of
the $E_{sym}(\rho)$? On the other hand, suppose one can not
distinguish different orientations in U+U collisions in an
unfortunate situation, how big are the relative effects of nuclear
symmetry energy and nuclear deformation/orientation on
isospin-sensitive observables (isospin tracers) in inclusive
reaction events? To help answer these questions, we examine relative
effects of the collision geometry and nuclear symmetry energy in U+U
collisions at 0.52 GeV/nucleon. Several $E_{sym}(\rho)$-sensitive
experimental observables in either tip-tip or body-body U+U
collisions are identified. The results are expected to be useful for
planning the U+U experiments at intermediate energies.

\section{Brief model description}

Our study is carried out within the isospin-dependent
Boltzmann-Uehling-Uhlenbeck (IBUU) transport model~\cite{IBUU}. To
facility discussions of the IBUU results, we shall firstly describe
briefly the effective nuclear interaction used in this study. The
neutron and proton distributions in uranium nuclei are then
outlined.

\subsection{The isospin- and momentum-dependent interaction}
\label{MDI}

The momentum dependence of the nuclear interaction is important in
understanding not only dynamics in intermediate-energy heavy-ion
collisions~\cite{GBD87,PDG88,MDYI90} but also thermodynamical
properties of nuclear matter~\cite{Xu07a,Xu07b,Xu08}. In the present
study we use an isospin- and momentum-dependent
interaction~\cite{Das03}. In this model, the mean-field potential of
a nucleon with momentum $\vec{p}$ and isospin $\tau$ is written as
\begin{eqnarray}
U(\rho,\delta ,\vec{p},\tau ) &=&A_{u}(x)\frac{\rho _{-\tau }}{\rho _{0}}%
+A_{l}(x)\frac{\rho _{\tau }}{\rho _{0}}  \notag \\
&+&B(\frac{\rho }{\rho _{0}})^{\sigma }(1-x\delta ^{2})-8\tau x\frac{B}{%
\sigma +1}\frac{\rho ^{\sigma -1}}{\rho _{0}^{\sigma }}\delta \rho
_{-\tau }
\notag \\
&+&\frac{2C_{\tau ,\tau }}{\rho _{0}}\int d^{3}p^{\prime }\frac{f_{\tau }(%
\vec{p}^{\prime })}{1+(\vec{p}-\vec{p}^{\prime })^{2}/\Lambda ^{2}}
\notag \\
&+&\frac{2C_{\tau ,-\tau }}{\rho _{0}}\int d^{3}p^{\prime }\frac{f_{-\tau }(%
\vec{p}^{\prime })}{1+(\vec{p}-\vec{p}^{\prime })^{2}/\Lambda ^{2}}.
\label{MDIU}
\end{eqnarray}%
In the above, $\rho=\rho_n+\rho_p$ is the nucleon number density and
$\delta=(\rho_n-\rho_p)/\rho$ is the isospin asymmetry of the
nuclear medium, with $\rho_{n(p)}$ the neutron (proton) density,
$\tau =1/2$ ($-1/2$) for neutrons (protons), and $f(\vec{p})$ is the
local phase-space distribution function. The values of the
parameters $A_{u}(x),A_{l}(x),\sigma,B,C_{\tau ,\tau },C_{\tau
,-\tau }$ and $\Lambda $ can be found in Refs.~\cite{Das03,Che05},
and they lead to the binding energy $-16$ MeV and incompressibility
$212$ MeV for symmetric nuclear matter and symmetry energy $30.5$
MeV at saturation density $\rho_0=0.16$ fm$^{-3}$.

\begin{figure}[h]
\centerline{\includegraphics[scale=0.8]{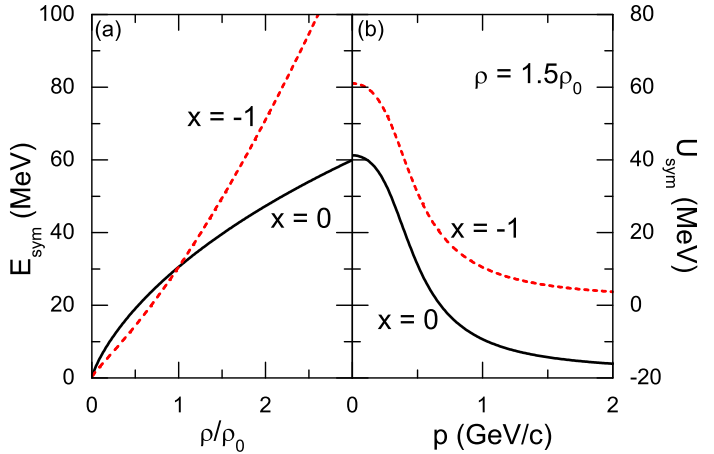}} \caption{(Color
online) The density dependence of the symmetry energy (a) and the
momentum dependence of the symmetry potential (b) at
$\rho=1.5\rho_0$. } \label{EUsym}
\end{figure}
The above mean-field potential comes from a modified Gogny force
including a zero-range effective three-body interaction and a
finite-range Yukawa-type two-body interaction~\cite{Das03,XuJ10},
where the parameter $x$ controls the relative contributions of the
isosinglet and isotriplet interaction channels. By adjusting the
value of $x$, the density dependence of the symmetry energy is
modified while the value of the symmetry energy at saturation
density and all the properties for symmetric nuclear matter remain
unchanged. The analytical formula of the symmetry energy in this
model can be found in Ref.~\cite{Xu09}. From comparing the IBUU
calculations with the isospin diffusion data, the value of $x$ is
constrained between $0$ and $-1$ at densities around and below the
saturation density of nuclear matter~\cite{Che05,Li05}. The
corresponding slope parameter $L=3\rho_0(dE_{sym}/d\rho)_{\rho_0}$
is between $L=60$ and $106$ MeV overlapping with constraints
extracted from studying several other observables using various
approaches (see \cite{Xu10} and references therein). At
supra-saturation densities, while the FOPI $\pi^-/\pi^+$ ratio data
favors IBUU calculations using a super soft symmetry energy with
$x=1$~\cite{Xia09}, the symmetry energy remains rather uncertain and
much more studies of additional observables are very much needed. To
evaluate the relative effects due to the collision geometry and the
symmetry energy on isospin tracers, we use $x=0$ and $-1$ in the
whole density range in the present work. We emphasize that the
current uncertainty range of the symmetry energy at supra-saturation
densities is much larger than the one considered here~\cite{LCK08}.
Thus, the symmetry energy effects with respect to those due to
geometry in U+U collisions studied here should be considered as
being rather conservative. As shown in the panel (a) of
Fig.~\ref{EUsym}, $x=-1$ leads to a stiffer symmetry energy which is
smaller at subsaturation densities and larger at supra-saturation
densities compared with $x=0$. A stiffer symmetry energy generally
leads to a larger isospin fractionation effect, i.e., a less
neutron-rich high-density regime and a more neutron-rich low-density
regime, to lower the energy of the whole system compared with a
softer symmetry energy.

The mean-field potential from Eq.~(\ref{MDIU}) can be approximately
expressed as
\begin{equation}
U_{n/p}(\rho,\delta,\vec{p}) \approx U_0(\rho,\vec{p}) \pm
U_{sym}(\rho,\vec{p}) \delta,
\end{equation}
where $U_0$ is the isoscalar mean-field potential, $U_{sym}$ is the
symmetry potential, and the $\pm$ sign is for neutrons (protons).
The momentum dependence of the symmetry potential at
$\rho=1.5\rho_0$ are plotted in the panel (b) of Fig.~\ref{EUsym}.
It is seen that the symmetry potential is mostly positive, which
means that generally neutrons feel a more repulsive potential than
protons. In addition, the symmetry potential is larger for $x=-1$ at
supra-saturation densities, thus neutrons are more likely to be
emitted from while protons are more likely to be trapped in the
high-density regime, resulting in a less neutron-rich high-density
regime and a more neutron-rich low-density regime, consistent with
the above discussion of the isospin fractionation effect. The
decreasing trend of the symmetry potential with increasing momentum
is consistent with the energy dependence of the Lane potential and
leads to larger effective masses of neutrons than
protons~\cite{Li04}. The in-medium nucleon-nucleon elastic cross
sections are consistently modified from their free-space values
according to the in-medium nucleon effective masses~\cite{Li05}.

\subsection{Initial density distributions}
\label{den}

Compared to the uniform ellipsoid distribution as used in
Refs.~\cite{Li00a,Luo07,Cao10}, a deformed Woods-Saxon distribution
is more realistic for uranium nuclei~\cite{Hag06}. In addition,
various studies~\cite{Bar95,Pom97,War98,Dob02,Sar07,Gai12} have
shown that neutron skin exists in deformed nuclei as well, and the
deformations of neutron and proton distributions are different. For
$^{238}U$, we use Woods-Saxon distributions for neutrons and protons
with different radii and deformations
\begin{equation}
\rho_{n/p}(r,\theta) =
\frac{\rho_{n/p}^0}{1+\exp[(r-R_{n/p}^\prime(\theta))/a]},
\end{equation}
where
\begin{eqnarray}
R_n^\prime(\theta) &=& R_n[1+\delta_n Y_2^0(\theta)], \label{rn}\\
R_p^\prime(\theta) &=& R_p[1+\delta_p Y_2^0(\theta)]. \label{rp}
\end{eqnarray}
In the above, $\rho_{n/p}^0$ is the normalization constant,
$Y_2^0(\theta)$ is the spherical harmonics, and $a=0.55$ fm is the
surface diffuseness parameter. In the present work we use $R_n=6.91$
fm for neutrons and $R_p=6.71$ fm for protons, respectively. The
deformation parameters are $\delta_n=0.275$ for neutrons and
$\delta_p=0.285$ for protons, and $R_n\delta_n \approx R_p\delta_p$
is approximately satisfied.

\section{Results and discussions}
\label{results}

\begin{figure}[h]
\includegraphics[scale=0.45]{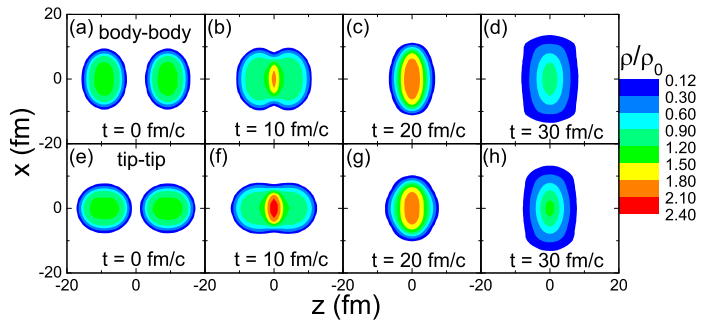}
\caption{(Color online) Time evolution of the density contours in
x-z plane for central body-body (upper panels) and tip-tip (lower
panels) U+U collisions.} \label{dencon}
\end{figure}

Let's first have a global picture of the body-body and tip-tip U+U
collisions in IBUU transport model, as shown in Fig.~\ref{dencon}.
Initially two uranium nuclei are placed at a distance along the
z-axis with different orientations. As time goes on, collision
happens and the central density can reach $2\rho_0$ or higher. The
system is most compressed around 20 fm/c, when the interaction is
the strongest and $\Delta$ resonances are abundantly produced. After
20 fm/c the system begins to expand and the density drops. One can
expect the different dynamic processes in body-body and tip-tip U+U
collisions from their different density evolutions, especially the
longer duration time of the high-density regime in tip-tip
collisions.

The neutron-proton differential transverse flow, different elliptic
flows of neutrons and protons, pre-equilibrium neutron/proton ratio,
and $\pi^-/\pi^+$ ratio are among the experimental observables known
to be sensitive to the $E_{sym}(\rho)$~\cite{LCK08,Rus11}. In the
following subsections, we will examine separately the sensitivity of
these observables to the variation of $E_{sym}(\rho)$ in both
tip-tip and body-body collisions at beam energy of 0.52 GeV/nucleon.
We will also compare effects of the symmetry energy with those due
to the collision geometry. Pre-equilibrium nucleons are identified
as those becoming 'free' (nucleons with local densities
$\rho<\rho_0/8$) earlier than $40$ fm/c when the participant is well
equilibrated. We found that our conclusions are not sensitive to
different criteria of identifying pre-equilibrium nucleons. Besides
the extreme collision geometries of body-body and tip-tip
collisions, as a reference we also study sphere-sphere collisions by
setting $\delta_n=0$ and $\delta_p=0$ in Eqs.~(\ref{rn}) and
(\ref{rp}). Although it is still very challenging, in the present
study we assume that desired reaction orientations can be achieved
by either initially polarizing the colliding nuclei or finally
selecting special events using orientation triggers proposed in the
literature. To simulate typical cases of central and
non-central collisions, we generated events with impact parameters $b=0$ fm and
$b=b_{\rm max}/2$ where $b_{\rm max}$ is about $12.4$, $13.6$, and
$16.0$ fm for tip-tip, sphere-sphere, and body-body collisions,
respectively. In each case about $200,000$ events are generated for
analysis.

\subsection{Neutron-proton differential transverse flow}

\begin{figure}[h]
\centerline{\includegraphics[scale=0.8]{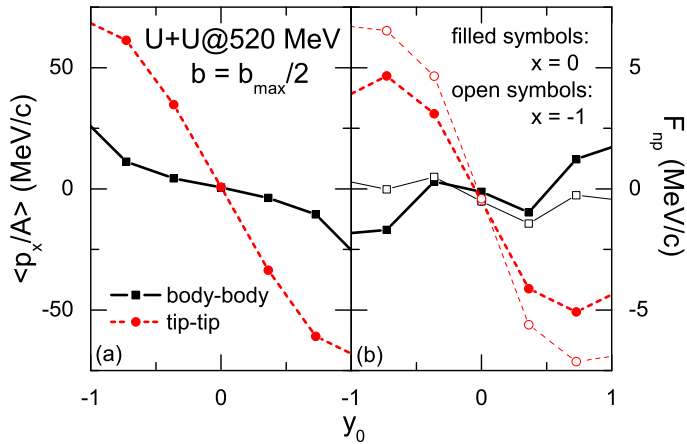}} \caption{(Color
online) Transverse flow (a) and neutron-proton differential
transverse flow (b) from non-central U+U collisions at beam energy
0.52 GeV/nucleon with different collision geometries and symmetry
energies.} \label{v1}
\end{figure}

It is well known that the transverse flow is a useful probe of the
nuclear EOS. Generally, since the isovector potential is relatively
small compared to the isoscalar potential, the transverse flow is
dominated by effects of the EOS of symmetric nuclear matter and
nucleon-nucleon scatterings.  The transverse flow in non-central U+U
collisions as a function of reduced rapidity $y_0=y/y_{\rm
beam}^{}$, where $y_{\rm beam}^{}$ is the beam rapidity in the
center-of-mass frame, is shown in the panel (a) of Fig.~\ref{v1}.
Similar to the finding in Ref.~\cite{Luo07}, the transverse flow is
found to be much larger in tip-tip collisions than in body-body
collisions due to the longer duration time of the high-density
phase, which leads to a more persisting pressure, and stronger
participant squeeze-out and spectator bounce-off
effects~\cite{Gut84} due to the collision geometry in the former
case.

To probe the isovector potential and thus the symmetry energy, it is
more useful to use the neutron-proton differential transverse flow
which minimizes effects of the EOS of symmetric nuclear matter and
nucleon-nucleon scatterings but adds up constructively effects due
to the positive and negative symmetry potentials of neutrons and
protons, respectively~\cite{Li00b}. As shown in the panel (b) of
Fig.~\ref{v1}, a stiffer symmetry energy ($x=-1$) leads to a larger
neutron-proton differential flow, reflecting a larger difference of
neutron and proton transverse flow. It is shown that the
neutron-proton differential transverse flow is always larger in
tip-tip collisions than in body-body collisions, and it is even
larger from a soft symmetry energy in tip-tip collisions than that
from a stiff symmetry energy in body-body collisions. Thus, effects
of the collision geometry on neutron-proton differential transverse
flow are much larger than those of the symmetry energy. The slope
parameters $(dF_{np}/dy_0)_{y_0=0}$ in different cases are shown in
Tab.~\ref{tab1}, where results from sphere-sphere collisions are
also included for comparison. The slope parameter in tip-tip
collisions is more sensitive to the symmetry energy than that in the
other two cases, and this is understandable as the total transverse
flow in tip-tip collisions is the largest and is most sensitive to
the mean-field potential. The neutron-proton differential transverse
flow in tip-tip collisions of deformed nuclei is thus a better probe
to determine the symmetry energy at supra-saturation densities
compared with collisions of spherical nuclei.

\begin{table}[h]
\caption{{\protect\small Slope parameters (in MeV/c) of
neutron-proton differential transverse flow with different collision
geometries and symmetry energies in non-central U+U collisions.}}
\label{tab1}
\begin{tabular}{ccccccc}
\hline\hline
  &  $\frac{dF_{np}}{dy_0}(x=0)$  \quad & $\frac{dF_{np}}{dy_0}(x=-1)$ \\
\hline
$$ body-body (non-central) &  $-1.75 \pm 0.46$  & $-2.64 \pm 0.46$\\
$$ tip-tip (non-central)  &  $-9.92 \pm 0.49$  & $-14.10 \pm 0.48$\\
$$ sphere-sphere (non-central)  &  $-5.15 \pm 0.48$  & $-7.68 \pm 0.48$\\
\hline\hline
\end{tabular}%
\end{table}

\subsection{Elliptic flow}

\begin{figure}[h]
\centerline{\includegraphics[scale=0.8]{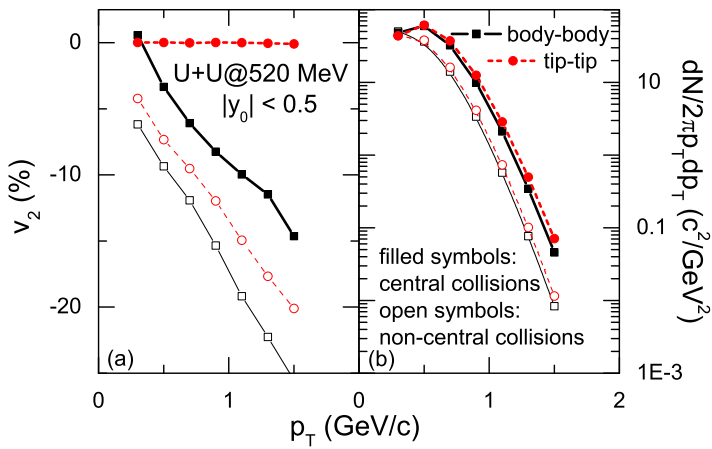}} \caption{(Color
online) Transverse momentum dependence of the elliptic flow (a) and
$p_T^{}$ spectra (b) of pre-equilibrium nucleons from central and
non-central U+U collisions at beam energy 0.52 GeV/nucleon with
different collision geometries. $x=0$ is used in the calculation.}
\label{pt}
\end{figure}

The elliptic flow has been used to study properties of hot and
dense matter formed in the early stage of heavy-ion collisions at both
relativistic and intermediate energies, see, e.g., Refs.~\cite{Son11} and~\cite{Dan02}.
To probe the high-density symmetry energy, we examine the transverse momentum ($p_T$) dependence
of elliptic flow ($v_2$) for mid-rapidity ($|y_0|<0.5$)
pre-equilibrium nucleons in central and non-central U+U collisions
from different collision geometries in the panel (a) of Fig.~\ref{pt}.
In non-central U+U collisions, $v_2$ is negative
for both tip-tip and body-body collisions, as the emission of
pre-equilibrium nucleons suffers the shadowing effects from the cold
spectators and they are mostly squeezed out in the direction perpendicular
to the reaction plane. Due to the geometry shape, body-body
collisions give a larger magnitude of $v_2$. The elliptic flow vanishes in
central tip-tip collisions due to symmetry as one expects. However, it is still considerable in central
body-body collisions. These are all consistent with previous results
in Refs.~\cite{Li00a,Luo07,Cao10}.

The elliptic flows of pre-equilibrium neutrons and protons are
affected differently by the symmetry potential too. The difference
of neutron and proton elliptic flows, especially at higher $p_T$, is
thus a useful probe of $E_{sym}(\rho)$ as well. As shown in
Tab.~\ref{tab2}, in central body-body U+U collisions, the expansion
of participants does not suffer from the shadowing effects, and a
more repulsive mean-field potential leads to a larger $v_2$. Due to
the more repulsive mean-field potential of neutrons than protons
considering the combined effects of symmetry potential, Coulomb
interaction, and initial eccentricity, the elliptic flow of neutrons
is slightly larger (more negative) than protons for a soft symmetry
energy ($x=0$), while the difference is larger for a stiff symmetry
energy ($x=-1$). In non-central collisions the situation is a little
different. As the expansion of the participant nucleons is blocked
by the spectator nucleons, a more repulsive potential leads to a
faster expansion of the participant matter and more nucleons are
squeezed out perpendicular to the reaction plane, resulting in a
more negative elliptic flow as discussed in
Refs.~\cite{Dan98,Dan02}. It is seen in Tab.~\ref{tab2} that the
elliptic flow of neutrons is somehow less negative than that of
protons, taking into account the combined effects of the symmetry
potential, Coulomb interaction, and initial eccentricity. Similar
results were obtained in Ref.~\cite{Rus11} from UrQMD calculations.
With a stiff symmetry energy, which leads to a more repulsive
neutron potential and a less repulsive proton potential at high
densities, the difference of high-$p_T$ neutron and proton elliptic
flow is smaller. It is seen that this difference is more sensitive
to the symmetry energy in body-body than tip-tip and sphere-sphere
collisions. This is again understandable as the total elliptic flow
is the largest in body-body collisions. Thus, the elliptic flow
difference between neutrons and protons in body-body U+U collisions
is a better probe of $E_{sym}(\rho)$ than collisions of spherical
nuclei of the same mass.

\begin{table}[h]
\caption{{\protect\small Mid-rapidity ($|y_0|<0.5$) high-$p_T$
($p_T>0.5$ GeV/c) neutron and proton elliptic flow difference ($\%$)
with different collision geometries and symmetry energies in central
and non-central U+U collisions.}} \label{tab2}
\begin{tabular}{ccccccc}
\hline\hline
  & $v_2(n)-v_2(p)$ \quad & $v_2(n)-v_2(p)$ \\
  & $(x=0)$ \quad & $(x=-1)$ \\
\hline
$$ body-body (central) & $-0.02 \pm 0.06$ & $-0.26 \pm 0.06$ \\
$$ body-body (non-central) & $1.46\pm 0.09$ & $1.03 \pm 0.09$ \\
$$ tip-tip (non-central) & $1.14\pm 0.08$ & $0.93 \pm 0.08$ \\
$$ sphere-sphere (non-central) & $1.21 \pm 0.09$ & $1.00 \pm 0.08$\\
\hline\hline
\end{tabular}%
\end{table}

\subsection{Pre-equilibrium neutron/proton ratio}

The transverse momentum spectrum is known to be sensitive not only
to the isoscalar potential and the nucleon-nucleon scattering cross
section, but also the collision geometries in reactions involving
deformed nuclei. As an example, the transverse momentum spectra of
mid-rapidity ($|y_0|<0.5$) pre-equilibrium nucleons from body-body
and tip-tip collisions are shown in the panel (b) of Fig.~\ref{pt}.
Similar to the finding in Ref.~\cite{Luo07}, the transverse momentum
spectrum is stiffer in tip-tip than body-body collisions, as the
high-density duration time is longer in tip-tip collisions and
high-$p_T$ nucleons are mainly emitted from the high-density regime.

It was first shown in Ref.~\cite{Li97} that the pre-equilibrium
neutron/proton (n/p) ratio, which is insensitive to the isoscalar
part of the nuclear EOS and the nucleon-nucleon scattering cross
section, is a sensitive probe of the density dependence of nuclear
symmetry energy. Due to the different $p_T$ spectra of
pre-equilibrium nucleons in body-body and tip-tip collisions, it is
reasonable to expect that the high-$p_T$ n/p ratio may depend on the
collision geometry as well. To test this expectation, the neutron/proton ratio of pre-equilibrium
nucleons from different cases are listed in Tab.~\ref{tab3}. The high-$p_T$ n/p ratios are larger with a stiff
symmetry energy ($x=-1$) due to the larger symmetry potential as
expected, and it is larger in non-central collisions than in central
collisions as a result of more neutron-rich participants in
non-central collisions. In addition, it is more sensitive to the
symmetry energy in central collisions than in non-central
collisions. In spite of the small symmetry energy effect in the
current collision energy, the n/p ratio is found to be slightly more
sensitive to the symmetry energy in tip-tip collisions than in the
other two cases. Moreover, the geometrical effect is rather small especially in
central collisions compared with the symmetry energy effect.

\begin{table}[h]
\caption{{\protect\small Mid-rapidity ($|y_0|<0.5$) high-$p_T$
($p_T>0.5$ GeV/c) neutron/proton ratios with different collision
geometries and symmetry energies in central and non-central U+U
collisions.}} \label{tab3}
\begin{tabular}{ccccccc}
\hline\hline
  & $N_n/N_p$ $(x=0)$ \quad & $N_n/N_p$ $(x=-1)$   \quad  \\
\hline
$$ body-body (central)  & $1.248 \pm 0.001$ & $1.283 \pm 0.001$ \\
$$ tip-tip (central)  & $1.241 \pm 0.001$ & $1.286 \pm 0.001$ \\
$$ sphere-sphere (central)  & $1.246 \pm 0.001$ & $1.284 \pm 0.001$ \\
$$ body-body (non-central)  & $1.292 \pm 0.002$ & $1.300 \pm 0.002$ \\
$$ tip-tip (non-central)  & $1.281 \pm 0.001$ & $1.295 \pm 0.001$ \\
$$ sphere-sphere (non-central)  & $1.290 \pm 0.002$ & $1.300\pm 0.002$ \\

\hline\hline
\end{tabular}%
\end{table}

\subsection{$\pi^-/\pi^+$ ratio}

\begin{figure}[h]
\centerline{\includegraphics[scale=0.8]{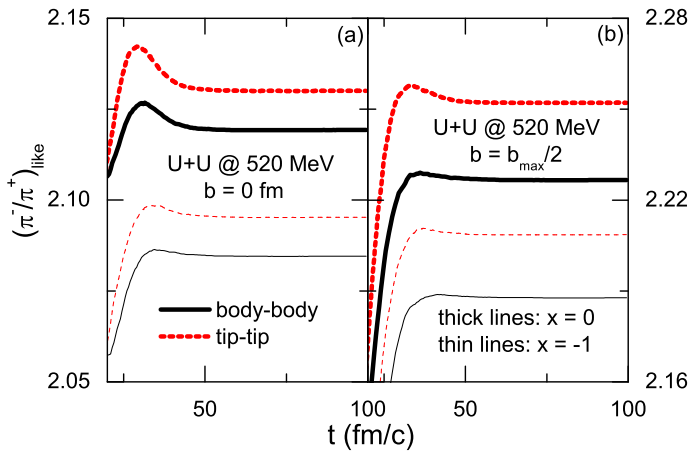}} \caption{(Color
online) Time evolution of the $(\pi^-/\pi^+)_{\rm like}$ ratio in
central (a) and non-central (b) U+U collisions with different
collision geometries and symmetry energies, in which the
contributions from $\Delta$ resonances are taken into
consideration.} \label{rpi2}
\end{figure}

The $\pi^-/\pi^+$ ratio is a sensitive tracer of the isospin
asymmetry of the high-density participants~\cite{Li02} which is
essentially determined by the $E_{sym}(\rho)$ through the isospin
fractionation process, as neutron-neutron (proton-proton) collisions
produce mostly $\Delta^-$ ($\Delta^{++}$) resonances which decay
into $\pi^-$ ($\pi^+$). A more systematic analysis has demonstrated
that the $\pi^-/\pi^+$ ratio is more sensitive to the symmetry
energy at supra-saturation densities but insensitive to that at
subsaturation densities~\cite{Zha09}. Figure~\ref{rpi2} displays the
time evolution of $(\pi^-/\pi^+)_{\rm like}$ ratio, in which the
contributions from $\Delta$ resonances are taken into account, from
different symmetry energies and collision geometries in both central
and non-central U+U collisions. A soft symmetry energy leads to a
larger $(\pi^-/\pi^+)_{\rm like}$ ratio as expected, although the
symmetry energy effects with $x=0$ and $x=-1$ are not large for the
collision energy studied here. However, it is interesting to note
that the resulting $(\pi^-/\pi^+)_{\rm like}$ ratio is higher in
tip-tip collisions. Effects of the collision geometry on the final
$\pi^-/\pi^+$ ratio are small but still around $30\%$ of those due
to the symmetry energy in central collisions, and they are even
larger in non-central collisions. This is mainly because pions are
strongly affected by the final state reabsorption and re-emission.
The $\pi^-/\pi^+$ ratio is thus affected significantly differently
by the different path-lengths in collisions of various geometries.
Nevertheless, we found that the sensitivities of the $\pi^-/\pi^+$
ratio to the symmetry energy are similar for different orientations.
To be more quantitative, values of the final $\pi^-/\pi^+$ ratio in
different cases are listed in Tab.~\ref{tab4}.

\begin{table}[h]
\caption{{\protect\small $\pi^-/\pi^+$ ratios with different
collision geometries and symmetry energies in central and
non-central U+U collisions.}} \label{tab4}
\begin{tabular}{ccccccc}
\hline\hline
  & $\pi^-/\pi^+$ $(x=0)$ \quad & $\pi^-/\pi^+$ $(x=-1)$   \quad  \\
\hline
$$ body-body (central)  & $2.119 \pm 0.002$ & $2.085 \pm 0.002$ \\
$$ tip-tip (central)  & $2.130 \pm 0.002$ & $2.095 \pm 0.002$ \\
$$ sphere-sphere (central)  & $2.123 \pm 0.002$ & $2.091 \pm 0.002$ \\
$$ body-body (non-central)  & $2.227 \pm 0.003$ & $2.188 \pm 0.003$ \\
$$ tip-tip (non-central)  & $2.252 \pm 0.003$ & $2.209 \pm 0.003$ \\
$$ sphere-sphere (non-central)  & $2.245 \pm 0.003$ & $2.197 \pm 0.003$ \\

\hline\hline
\end{tabular}%
\end{table}

\section{Summary}
\label{summary}

We studied the effects of the collision geometry on observables that
are known to be sensitive to the nuclear symmetry energy, such as
the neutron-proton differential transverse flow, the difference of
neutron and proton elliptic flow, the pre-equilibrium neutron/proton
ratio and the $\pi^-/\pi^+$ ratio in U+U collisions. We found that
the neutron-proton differential transverse flow is always larger in
tip-tip than body-body collisions, and it is more sensitive to the
symmetry energy in the former case. On the other hand, the
difference of neutron and proton elliptic flow is more sensitive to
the symmetry energy in body-body than tip-tip collisions. In
addition, the n/p ratio of pre-equilibrium nucleons is slightly more
sensitive to the symmetry energy in tip-tip collisions. Moreover,
the collision geometry is found to affect significantly the
$\pi^-/\pi^+$ ratio.

As the collision geometry effects are comparable to or sometimes
even larger than the symmetry energy effects, one should be careful
when studying isospin-sensitive observables in collisions involving
deformed nuclei. Large errors may be induced by assuming that the
colliding nuclei are spherical priorly. In addition, using deformed
nuclei, the neutron-proton differential transverse flow in tip-tip
collisions and the difference of neutron and proton elliptic flow in
body-body collisions are better probes for symmetry energy at
supra-saturation densities compared with collisions of spherical
nuclei.

\begin{large}
\textbf{Acknowledgements}
\end{large}

We thank Dr. Derek Harter and Dr. Sam Saffer for making available to
us facilities at the high-performance Computational Science Research
Cluster at Texas A$\&$M University-Commerce. Zachary Martinot would
like to thank the NSF funded REU (Research Experience for
Undergraduates) program at Texas A$\&$M University-Commerce for the
opportunity of participating in the research reported in this paper.
We would also like to thank Dr. W. G. Newton and Dr. F. J. Fattoyev
for helpful discussions. This work was supported in part by the US
National Science Foundation grants PHY-0757839, PHY-1068022 and REU
grant no. 1062613 as well as the National Aeronautics and Space
Administration under grant NNX11AC41G issued through the Science
Mission Directorate.


\begin{thebibliography}{99}

\bibitem{peter} P. Braun-Munzinger, Memorandum to RHIC management
on uranium beams in RHIC, Sept. 18, 1992.

\bibitem{ed99b} E. V. Shuryak, Phys. Rev. C \textbf{61}, 034905 (2000).

\bibitem{Kol00} P. F. Kolb, J. Sollfrank, and U. Heinz, Phys. Rev. C
\textbf{62}, 054909 (2000).

\bibitem{Hei05} U. Heinz and A. Kuhlman, Phys. Rev. Lett.
\textbf{94}, 132301 (2005).

\bibitem{Nep07} C. Nepali, G. Fai, and D. Keane, Phys. Rev. C \textbf{76}, 051902(R)
(2007).

\bibitem{Fil09} P. Filip, R. Lednicky, H. Masui, and N. Xu, Phys.
Rev. C \textbf{80}, 054903 (2009).

\bibitem{Hir11} T. Hirano, P. Huovinen, and Y. Nara, Phys. Rev. C
\textbf{83}, 021902(R) (2011).

\bibitem{Haq12} M. R. Haque, Z. W. Lin, and B. Mohanty, Phys.
Rev. C \textbf{85}, 034905 (2012).

\bibitem{Vol10} Sergei A. Voloshin, Phys. Rev. Lett. \textbf{105}, 172301 (2010).

\bibitem{Gag12} C. Gagliardi, Talk at the 11th International Conference on
Nucleus-Nucleus Collisions, San Antonio, Texas, USA, May 27-June 1,
2012. \\
\url{http://cyclotron.tamu.edu/nn2012/Slides/Plenary/RHIC_Results_NN2012_Gagliardi.pdf}

\bibitem{Luo07} X. F. Luo, X. Dong, M. Shao, K. J. Wu, C. Li, H. F. Chen, and H. S. Xu, Phys. Rev. C \textbf{76}, 044902 (2007).

\bibitem{Cao10} X. G. Cao, G. Q. Zhang, X. Z. Cai, Y. G. Ma, W. Guo, J. G. Chen, W. D. Tian, D. Q. Fang, and H. W.
Wang, Phys. Rev. C \textbf{81}, 061603(R) (2010).

\bibitem{Li00a} B. A. Li, Phys. Rev. C \textbf{61}, 021903(R) (2000).

\bibitem{Wu08} K. J. Wu, F. Liu and N. Xu, arXiv:0811.3044v1 [nucl-th]

\bibitem{Xu} N. Xu, private communications.

\bibitem{Xiao} Z. G. Xiao, private communications.

\bibitem{Ste05} A. W. Steiner, M. Prakash, J. M. Lattimer, and P. J. Ellis, Phys.
Rep. \textbf{411}, 325 (2005).

\bibitem{Lat07} J. M. Lattimer and M. Prakash, Science \textbf{304}, 536 (2004); Phys.
Rep. \textbf{442}, 109 (2007).

\bibitem{LCK08} B. A. Li, L. W. Chen, and C. M. Ko, Phys. Rep. \textbf{464},
113 (2008).

\bibitem{Lat01} J. M. Lattimer and M. Prakash, Astrophys. J. \textbf{550}, 426 (2001).

\bibitem{IBUU} B. A. Li, C. B. Das, S. Das Gupta and C. Gale, Phys. Rev. C \textbf{69},
011603(R) (2004); Nucl. Phys. A \textbf{735}, 563 (2004).

\bibitem{GBD87} C. Gale, G. Bertsch, and S. Das Gupta, Phys. Rev. C
\textbf{35}, 1666 (1987).

\bibitem{PDG88} G. M. Welke, M. Prakash, T. T. S. Kuo, S. Das Gupta,
and C. Gale, Phys. Rev. C \textbf{38}, 2101 (1988).

\bibitem{MDYI90} C. Gale, G. M. Welke, M. Prakash, S. J. Lee, and S. Das Gupta, Phys. Rev. C \textbf{41}, 1545 (1990).

\bibitem{Xu07a} J. Xu, L. W. Chen, B. A. Li, and H. R. Ma,
Phys. Rev. C \textbf{75}, 014607 (2007).

\bibitem{Xu07b} J. Xu, L. W. Chen, B. A. Li, and H. R. Ma,
Phys. Lett. \textbf{B650}, 348 (2007).

\bibitem{Xu08} J. Xu, L. W. Chen, B. A. Li, and H. R. Ma,
Phys. Rev. C \textbf{77}, 014302 (2008).

\bibitem{Das03} C. B. Das, S. Das Gupta, C. Gale, and B. A. Li, Phys. Rev. C \textbf{67}, 034611 (2003).

\bibitem{Che05} L. W. Chen, C. M. Ko, and B. A. Li, Phys. Rev. Lett. \textbf{94}, 032701
(2005).

\bibitem{XuJ10} J. Xu and C. M. Ko, Phys. Rev. C \textbf{82}, 044311
(2010).

\bibitem{Xu09} J. Xu, L. W. Chen, B. A. Li, and H. R. Ma, Astrophys. J. \textbf{697}, 1549 (2009).

\bibitem{Li05} B. A. Li and L. W. Chen, Phys. Rev. C \textbf{72},
064611 (2005).

\bibitem{Xu10} C. Xu, B. A. Li, and L. W. Chen, Phys. Rev. C \textbf{82}, 054607 (2010).

\bibitem{Xia09} Z. G. Xiao, B. A. Li, L. W. Chen, G. C. Yong, and M.
Zhang, Phys. Rev. Lett. \textbf{102}, 062502 (2009).

\bibitem{Li04} B. A. Li, Phys. Rev. C \textbf{69}, 064602 (2004).

\bibitem{Hag06} K. Hagino, N. W. Lwin, and M. Yamagami, Phys. Rev. C
\textbf{74}, 017310 (2006).

\bibitem{Bar95} A. Baran, J. L. Egido, B. Nerlo-Pomorska, K. Pomorski, P.
Ring, and L. M. Robledo, J. Phys. G \textbf{21}, 657 (1995).

\bibitem{Pom97} K. Pomorski, P. Ring, G. A. Lalazissis, A. Baran, Z. Lojewski,
B. Nerlo-Pomorska, M. Warda, Nucl. Phys. \textbf{A624}, 349 (1997).

\bibitem{War98} M. Warda, B. Nerlo-Pomorska, K. Pomorski, Nucl.
Phys. \textbf{A635}, 484 (1998).

\bibitem{Dob02} A. Dobrowolski, K. Pomorski, and J. Bartel, Phys.
Rev. C \textbf{65}, 041306(R) (2002).

\bibitem{Sar07} P. Sarriguren, M. K. Gaidarov, E. Moya de Guerra, and A. N.
Antonov, Phys. Rev. C \textbf{76}, 044322 (2007).

\bibitem{Gai12} M. K. Gaidarov, A. N. Antonov, P. Sarriguren, and E. Moya de
Guerra, Phys. Rev. C \textbf{85}, 064319 (2012).

\bibitem{Rus11} P. Russotto, P. Z. Wu, M. Zoricc, M. Chartier, Y. Leifels, R. C. Lemmon, Q. Li, J. {\L}ukasik,
A. Pagano, P. Paw{\l}owskig, W. Trautmann, Phys. Lett.
\textbf{B697}, 471 (2011).

\bibitem{Gut84} H. A. Gustafsson, {\it et al.}, Phys. Rev. Lett. \textbf{52}, 1590 (1984).

\bibitem{Li00b} B. A. Li, Phys. Rev. Lett. \textbf{85}, 004221
(2000).

\bibitem{Son11} H. C. Song, S. A. Bass, U. Heinz, T. Hirano, and C.
Shen, Phys. Rev. Lett. \textbf{106}, 192301 (2011).

\bibitem{Dan02} P. Danielewicz, R. Lacey, W. G. Lynch, Sceince
\textbf{298}, 1592 (2002).

\bibitem{Dan98} P. Danielewicz, Roy A. Lacey, P. B. Gossiaux, C. Pinkenburg,
P. Chung, J. M. Alexander, and R. L. McGrath, Phys. Rev. Lett.
\textbf{81}, 2438 (1998).

\bibitem{Li97} B. A. Li, C. M. Ko, and Z. Z. Ren, Phys. Rev. Lett.
\textbf{78}, 1644 (1997).

\bibitem{Li02} B. A. Li, Phys. Rev. Lett. \textbf{88}, 192701 (2002);
Nucl. Phys. \textbf{A708}, 365 (2002).

\bibitem{Zha09} M. Zhang, Z. G. Xiao, B. A. Li, L. W. Chen, G. C. Yong, S.
J. Zhu, Phys. Rev. C \textbf{80}, 034616 (2009).

\end{thebibliography}
\end{document}